\documentclass[10pt,letterpaper]{article}
\usepackage[top=0.85in,left=2.75in,footskip=0.75in,marginparwidth=2in]{geometry}

\usepackage[utf8]{inputenc}

\usepackage{cite}
\usepackage{lscape}

\usepackage{nameref,hyperref}

\usepackage[right]{lineno}

\usepackage{microtype}
\DisableLigatures[f]{encoding = *, family = * }

\raggedright
\usepackage{changepage}

\usepackage[aboveskip=1pt,labelfont=bf,labelsep=period,singlelinecheck=off]{caption}

\makeatletter
\renewcommand{\@biblabel}[1]{\quad#1.}
\makeatother

\usepackage{lastpage,fancyhdr,graphicx}
\usepackage{epstopdf}
\fancyheadoffset[L]{2.25in}
\fancyfootoffset[L]{2.25in}

\usepackage{color}

\definecolor{Gray}{gray}{.25}

\usepackage{graphicx}

\usepackage{sidecap}

\usepackage{wrapfig}
\usepackage[pscoord]{eso-pic}
\usepackage[fulladjust]{marginnote}
\reversemarginpar

\begin{document}
\vspace*{0.35in}

\begin{flushleft}
{\Large
\textbf\newline{Performance Comparison of Deep Learning Architectures 
\\for Artifact Removal in Gastrointestinal Endoscopic Imaging}
}
\newline
\\

Taira Watanabe\textsuperscript{1},
Kensuke Tanioka\textsuperscript{2},
Satoru Hiwa\textsuperscript{2},
Tomoyuki Hiroyasu\textsuperscript{2,*}
\\
\bigskip
\bf{1} Graduate school of Life and Medical Sciences, Doshisha University, Japan
\\
\bf{2} Department of Biomedical Sciences and Informatics, Doshisha University, Japan
\\
\bf{3} AI x Humanity Research Center, Doshisha University, Japan
\\
\bigskip
* tomo@is.doshisha.ac.jp

\end{flushleft}

\section*{Abstract}
Endoscopic images typically contain several artifacts. The artifacts significantly impact image analysis result in computer-aided diagnosis. Convolutional neural networks (CNNs), a type of deep learning, can removes such artifacts. Various architectures have been proposed for the CNNs, and the accuracy of artifact removal varies depending on the choice of architecture. Therefore, it is necessary to determine the artifact removal accuracy, depending on the selected architecture. In this study, we focus on endoscopic surgical instruments as artifacts, and determine and discuss the artifact removal accuracy using seven different CNN architectures.

\section*{keyword}
Deep Learning, Model Architecture, Convolutional Neural Networks, Artifact, Semantic Segmentation, Encoder-Decoder Network

\section{Introduction}
In recent years, artificial intelligence (AI) and data science technologies have been extensively employed in the field of medical endoscopic imaging owing to the improved performance of computing hardware, such as GPUs, and large-scale big data analysis technologies.
Research and development of computer-aided diagnosis (CADx) systems, which analyze endoscope images using supervised machine learning methods such as deep learning (DL) to assist doctors in diagnosis, is being promoted worldwide.
One such example is a system that automatically detects cancerous regions and polyps from the captured endoscopic images\cite{cancer2018seg, jha2021polyp}.

In most cases, endoscopic images contain artifacts, such as specular reflections, air bubbles, and surgical instruments.
Many of the CADx systems proposed in recent years analyze endoscopic images containing artifacts and use them for diagnosis.
However, artifacts have been reported to make tissue visualization challenging and significantly hinder quantitative image analysis. Simultaneously, artifacts may cause misinterpretation of CADx systems\cite{ali2020objective}. 
In particular, when supervised learning, such as DL, which has been used extensively in recent years, is used, the problem is that during the training phase, a learner may use artifacts for
identification instead of tissue.
Therefore, it is essential to improve the identification accuracy of the learner and remove the non-target information contained in the training data. It follows that, in an ideal CADx system, artifacts must be removed from the image as a preprocessing step during image input.

One of the most effective methods for artifacts removal is convolutional neural networks (CNNs), a type of DL.
The CNNs are primarily used for image recognition; various architectures have been proposed.
In the ImageNet Large Scale Visual Recognition Challenge (ILSVRC), various high-accuracy CNN architectures were proposed from 2012 to 2017; they are considered to have surpassed human recognition performance in 2015.
These architectures are used as de facto standard models to solve various problems, including image recognition. 
In addition to image recognition, the CNNs have also been used to solve object detection and segmentation tasks.
The CADx system mentioned above use semantic segmentation to automatically detect cancerous regions and polyps.
The semantic segmentation is a method for classifying each pixel of an image into instances and associating a class or category with the instances.
The semantic segmentation with CNNs starts with a fully convolutional network (FCN)\cite{long2015FCN}, which omits the fully connected layer of the conventional CNNs.
Subsequently, SegNet\cite{badrinarayanan2017segnet}, an encoder-decoder network consisting of an encoder that extracts features of the input image and a decoder that segments the extracted features, was proposed.
Similar to SegNet, U-Net\cite{u-net2015} is an encoder-decoder network with additional skip connections, is a CNN architecture often used for image segmentation.

To perform semantic segmentation, a model architecture needs to be selected.
The accuracy of artifact removal varies depending on the selected architecture.
Therefore, the accuracy of each CNN architecture for removing specific artifacts from endoscopic images must be clarified.

This study examines the accuracy of a model built with a CNN architecture when surgical instruments appear as artifacts in endoscopic images.
Herein, we segmented surgical instruments as artifacts using the Kvasir-Instrument dataset, which includes diagnostic and therapeutic surgical instruments in images of various conditions obtained through gastrointestinal endoscopy.
The underlying architecture was U-Net, which is an encoder-decoder network.
It has been reported that fine-tuning of the pre-trained encoder of U-Net is effective\cite{iglovikov2018ternausnet}.
Therefore, we changed the encoder of U-Net to seven CNN architectures and prepared seven different model architectures. For implementing the U-Net, we used "Segmentation Models PyTorch"\cite{Yakubovskiy:2019},
which is an open-source Python library for image segmentation.

Furthermore, we used the segmentation results presented in a previous study\cite{jha2021kvasir} as a baseline using simple U-Net and DoubleU-Net\cite{jha2020doubleu}.
In the Discussion, we compared the segmentation results of the seven networks with modified architectures with simple U-Net and DoubleU-Net.

\section{Materials and Methods}
\subsection{Dataset}
In this study, we used the Kvasir-Instrument dataset [9], which is an open-source dataset of gastrointestinal endoscopic images and is used for assessing disease burden and pathological resection volume.
The images contain surgical instruments for diagnostic and therapeutic purposes in analytical gastrointestinal endoscopy. The dataset can be downloaded from \url{https://datasets.simula.no/kvasir-instrument/}. 
All images were collected using standard Olympus and Pentax endoscopic equipment.
These images contain surgical instruments such as snares, balloons, and biopsy forceps.

\begin{figure}[htbt]
  \includegraphics[width=0.50\linewidth]{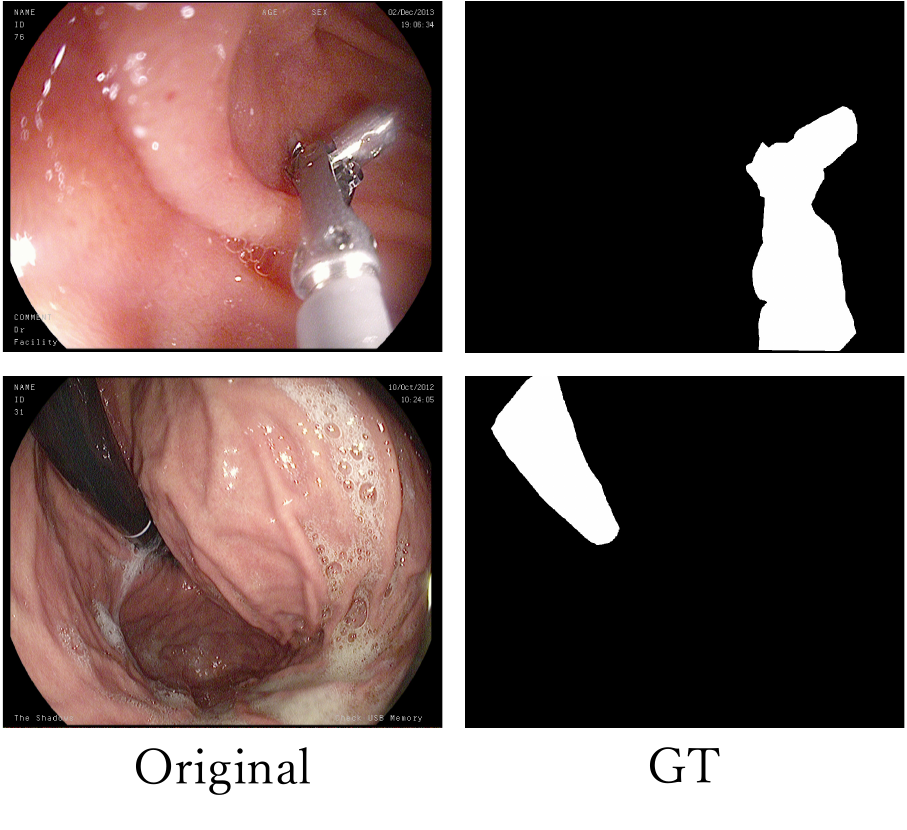}
 \caption{Sample images from Kvasir-Instrument}
 \label{fig:Sample images from Kvasir-Instrument}
 \end{figure}

The dataset consists of 590 endoscopic and pixel-level ground-truth (GT) mask images and bounding box information corresponding to the surgical instruments in the images.
In this study, we used the endoscope and the GT images. 
Figure \ref{fig:Sample images from Kvasir-Instrument} shows the sample and GT images.

The resolution of the images ranged from 523$\times$613 px to 1072$\times$1920 px.
In the present study, simple U-Net and DoubleU-Net have been constructed.
Information on which images are training and testing images is provided.
This study used images for training and testing according to this information.

\subsection{Experimental setup and implementation details}
In this study, we used U-Net\cite{u-net2015}, the most famous CNN architecture for image segmentation, as the basic architecture.
The U-Net is an encoder-decoder model consisting of an encoder that extracts features from the input image and a decoder that performs segmentation on the extracted features.
We changed the encoder of U-Net to one of seven prepared architectures.
The seven prepared architectures were ResNet-34\cite{he2016resnet}, DenseNet-121\cite{huang2017densenet}, SE-ResNet-50\cite{hu2018senet}, VGG-19\cite{simonyan2014VGG-19}, EfficientNet-B0\cite{tan2019efficientnet}, Inception-ResNet-v2\cite{szegedy2017inception}, and MobileNetV2\cite{sandler2018mobilenetv2}. 
All implementations were performed using PyTorch. 
The GPU used was an NVIDIA Tesla K80.

\begin{figure*}[htbt]
 \begin{center}
  \includegraphics[width=0.90\linewidth]{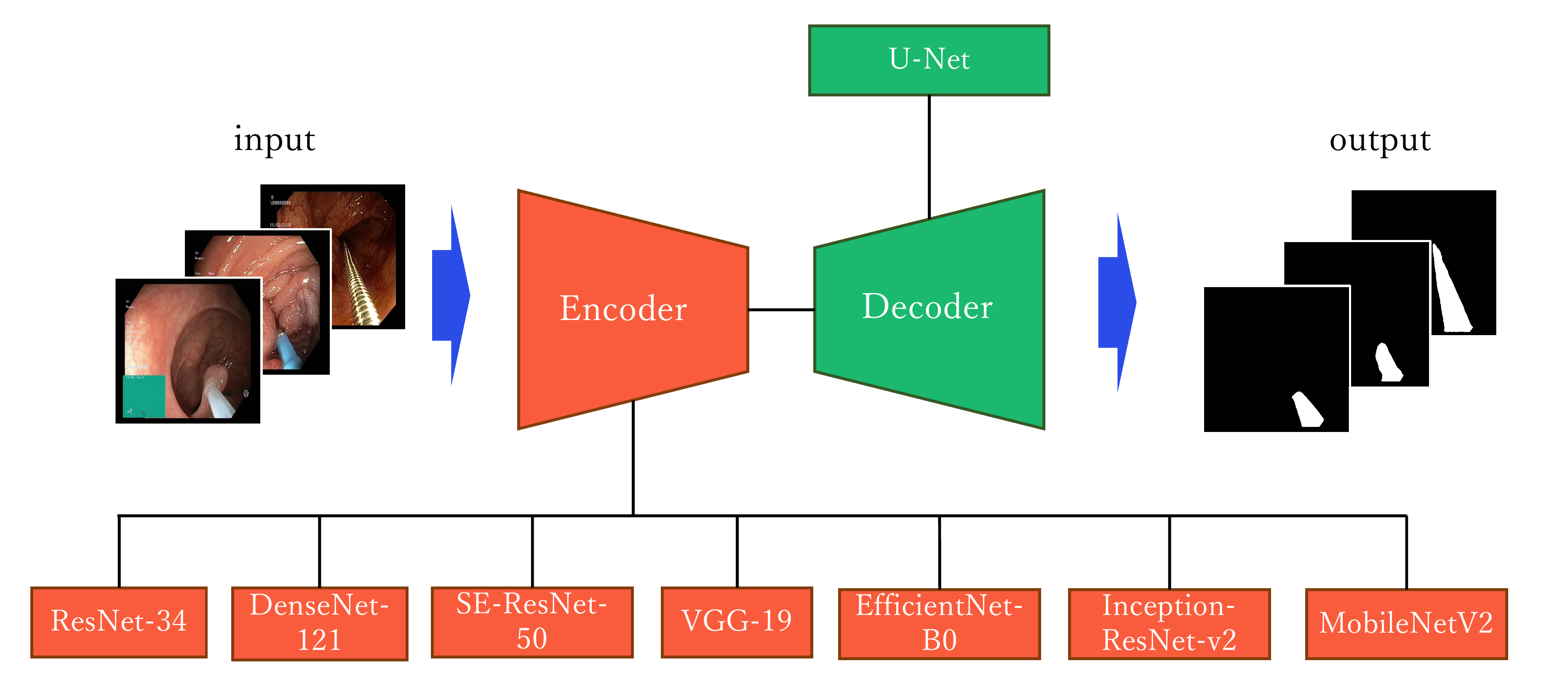}
 \end{center}
 \vspace{-5mm}
 \caption{Experimental Ditails}
 \label{fig:Experiment Ditails}
 \end{figure*}

The implementation of U-Net is based on "Segmentation Models Pytorch"\cite{Yakubovskiy:2019}, which is an open-source Python library for image segmentation.
“Segmentation Models Pytorch" has been used in many existing studies on segmentation tasks\cite{ziller2021medical, folle2021deep, xia2021benchmark}.
There are also several encoder architectures for the "Segmentation Models Pytorch," all of which use pre-trained models with initial network weights learned from ImageNet images.
Using this Kvasir- Instrument dataset and "Segmentation Models Pytorch," we fine-tuned the network by combining each architecture in the encoder of U-Net.
An overview of the experiment is shown in Figure \ref{fig:Experiment Ditails}.

 \vspace{0.01mm}
 \vspace{0.01mm}
 
The dataset was divided into training and test sets with distributions of 80\% (n = 472) and 20\% (n = 118), respectively.
These classifications are identical to those of the existing studies.
Furthermore, 20\% (n=95) of the training sets were randomly divided into validation sets.
Finally, 377, 95, and 118 cards were assigned to the training, validation, and test sets, respectively.
The training set was used for learning, and the validation set was used for selecting the optimal learner from several candidate learners constructed using the training set; the best learner was the one whose loss function had the smallest value after an epoch of the validation set.
The test set was used to evaluate the performance of the learner selected by the validation set.
 
Because each image in the dataset had different resolution, the images were resized to 512$\times$512 px to fit the input of U-Net.
Furthermore, data augmentation was used to increase the number of images in the dataset and prevent over-fitting.
Three primary data augmentation techniques, horizontal flip, vertical flip, and random rotation, were applied.
Adam optimizer with $\beta_1$=0.9 and $\beta_2$=0.999 was used for all networks; training was conducted for 40 epochs.
The batch size was eight, and the initial learning rate was 0.0001. After 25 epochs, the learning rate reduced to 0.00001.
The loss function for network training was the Dice loss function, a commonly used loss function in segmentation focused CNNs, is based on the Dice coefficient described in the evaluation metrics section.

The segmentation results obtained by the learner for each of the seven architectures were compared with the results of the basic architectures (i.e., U-Net and DoubleU-Net) used in the previous study\cite{jha2021kvasir}.

 \vspace{-4mm}
\subsection{Evaluation Metricss}
To quantitatively evaluate the segmentation results, the Jaccard similarity coefficient (JSC), Dice similarity coefficient (DSC), precision, recall, and accuracy are commonly used metrics in computer vision.
These metrics were calculated using the confusion matrix shown in Table \ref{Confusion matrix for a two class problem}. 

\begin{table}[htb]
\centering 
\caption{Confusion matrix for a two class problem}
\label{Confusion matrix for a two class problem}
\small 
\begin{tabular*}{\linewidth}{c|c|c}
\hline 
\multicolumn{1}{c|}{}
& \multicolumn{1}{c|}{\begin{tabular}{c} Predicted Class\\ (Positive Class)\end{tabular} }
& \multicolumn{1}{c}{\begin{tabular}{c} Predicted Class\\ (Negative Class)\end{tabular} }\\
\hline 
\hline 
\begin{tabular}{c} Actual Class\\ (Positive Class)\end{tabular} & True Positive & False Negative\\
\hline 
\begin{tabular}{c} Actual Class\\ (Negative Class)\end{tabular} & False Posituve & True Negative\\
\hline 
\end{tabular*}
\end{table}
\vspace{4mm}

 The number of pixels that were determined to be surgical instruments by the segmentation network and were also determined to be surgical instruments in GT was defined as true positive (TP), and the number of pixels that were not determined to be surgical instruments in GT was defined as false positive (FP). The number of pixels considered surgical instruments in GT was defined as false negative (FN), and the number of pixels that were not considered to be surgical instruments in GT was defined as true negative (TN).

Based on the values of TP, FP, TN, and FN obtained from the confusion matrix, the values of JSC, DSC, precision, recall, and accuracy were calculated using the following equations:

\begin{equation}
   Jaccard\; Coefficient =  \frac{TP}{TP + FP + FN}
   \label{Eq:one}
\end{equation}

 \begin{equation}
   DSC = \frac{2 \cdot TP}{2 \cdot TP + FP + FN}
   \label{equ:two}
\end{equation}

\begin{equation}
	Precision=\frac{TP}{TP+FP}
	\label{equ:three}
\end{equation}

\begin{equation}
	Recall=\frac{TP}{TP+FN}
  \label{equ:four}
\end{equation}

\begin{equation}
	Accuracy=\frac{TP+TN}{TP+FN+FP+TN}
	\label{equ:five}
\end{equation}
 \vspace{1mm}

The DSC calculated from Equation 2 is used in the loss function (i.e., Dice loss) expressed in Equation 6.

\begin{equation}
  Dice_{loss} = 1 - DSC
  \label{Eq:six}
\end{equation}

\section{Results}
\subsection{Quantitative evaluation}
For simple U-Net and DoubleU-Net used in previous studies, the values of JSCs obtained from segmentation results for Kvasir-Instrument were 85.8\% and 84.3\%, respectively \cite{jha2021kvasir}.

Table \ref{Comparison of results for each method (U-Net)} lists the values of the evaluation metrics obtained from the segmentation results of each network that uses a CNN architecture as an encoder of U-Net.

All the seven encoder architectures used in this experiment exceeded the values of existing studies in terms of JSC, DSC, precision, recall, and accuracy.
Among them, the network that used Inception-ResNet-v2 showed the highest JSC, DSC, precision, and accuracy values at 91.7\%, 95.0\%, 95.4\%, and 99.1\%, respectively.
As for recall, the network that used ResNet-34 showed the highest value at 96.3\%.

\subsection{Qualitative evaluation}
\begin{figure}[htb]
 \begin{center}
  \includegraphics[width=\linewidth]{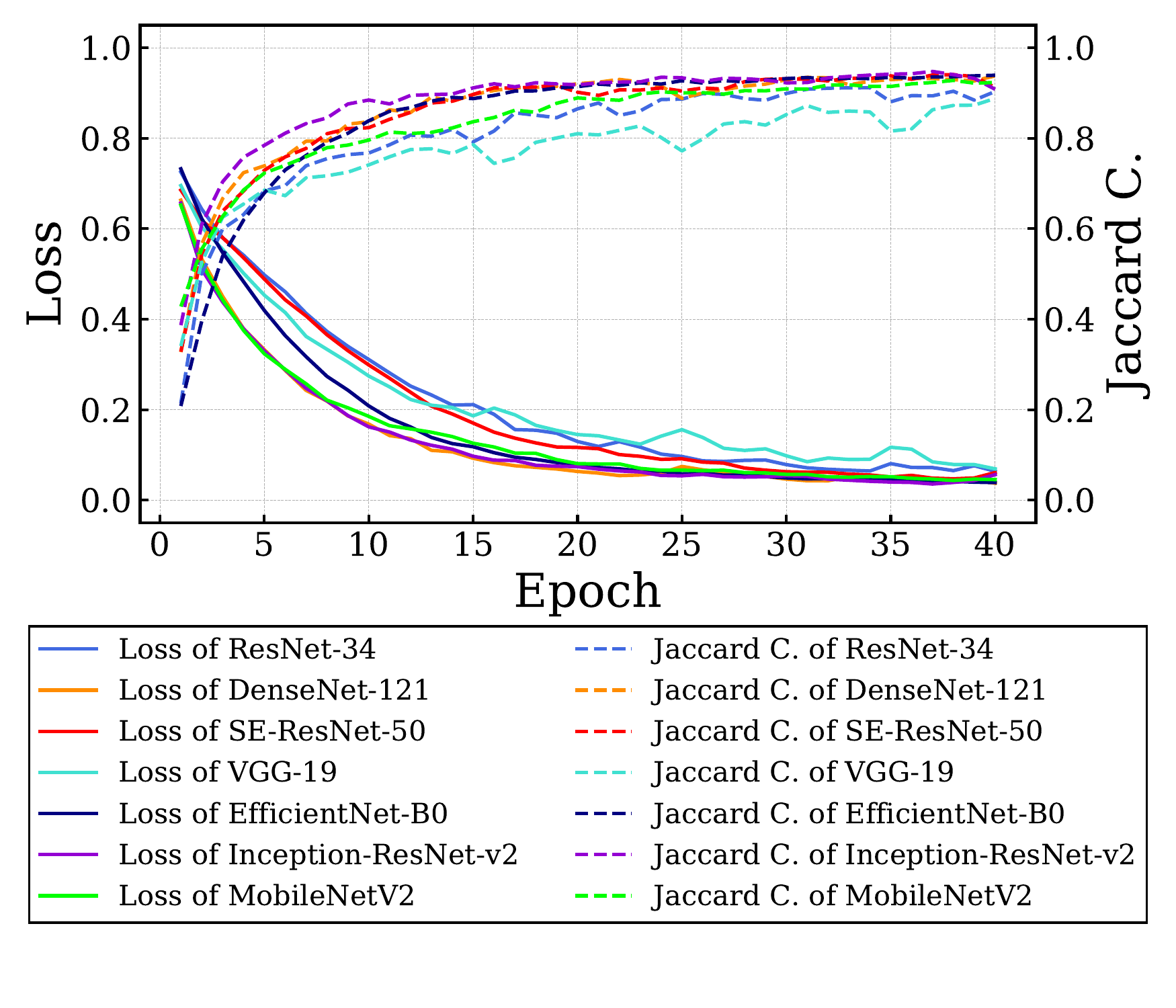}
 \end{center}
 \vspace{-12mm}
 \caption{Study curve of training}
 \label{fig:Study curve of training}
 \end{figure}

\begin{figure}[htb]
 \begin{center}
  \includegraphics[width=\linewidth]{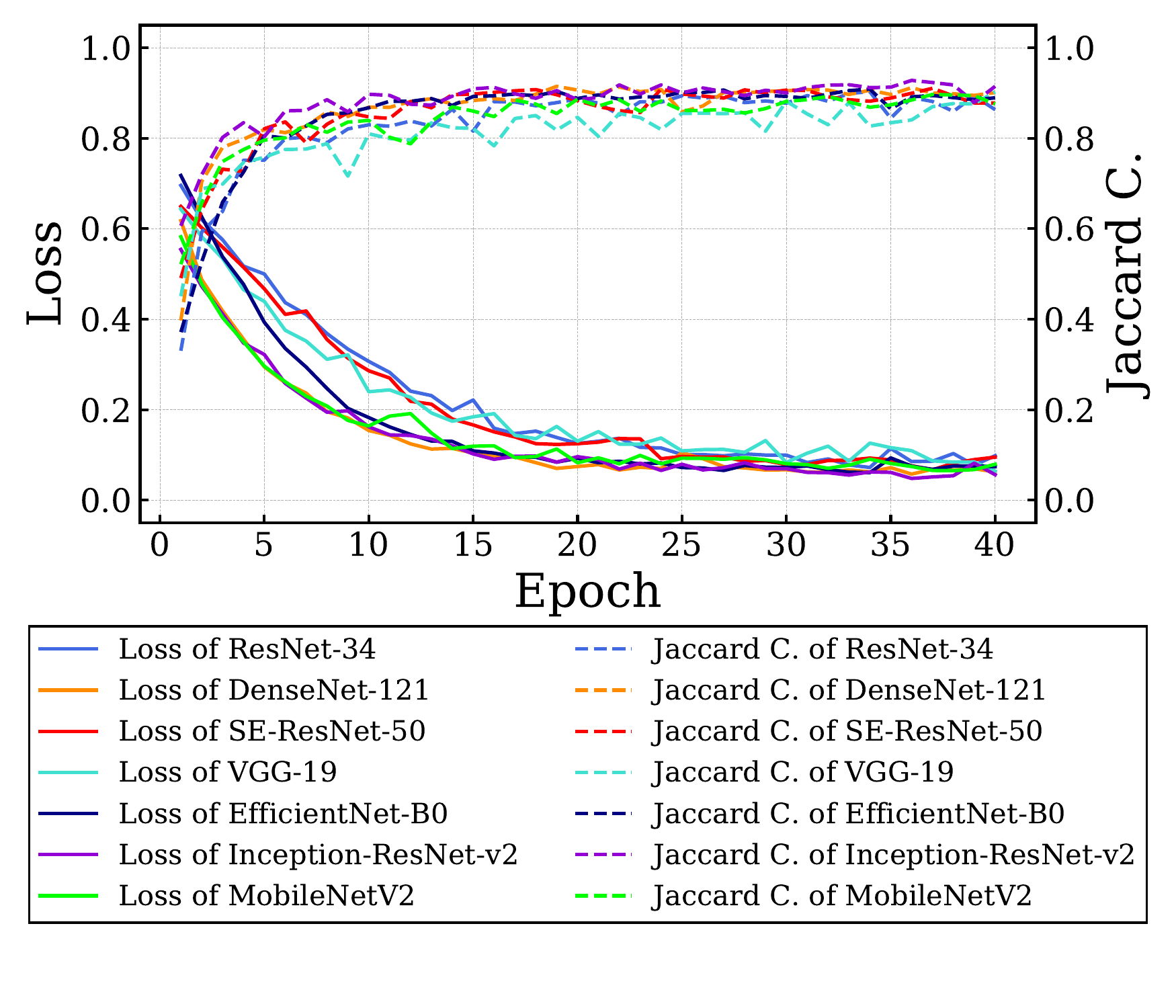}
 \end{center}
 \vspace{-12mm}
 \caption{Study curve of validation}
 \label{fig:Study curve of validation}
 \end{figure}

The Dice loss and JSCs of each CNN architecture at each epoch for the training and validation sets are shown in Figure \ref{fig:Study curve of training} and Figure \ref{fig:Study curve of validation}, respectively.

\begin{figure}[htb]
\includegraphics[width=\columnwidth]{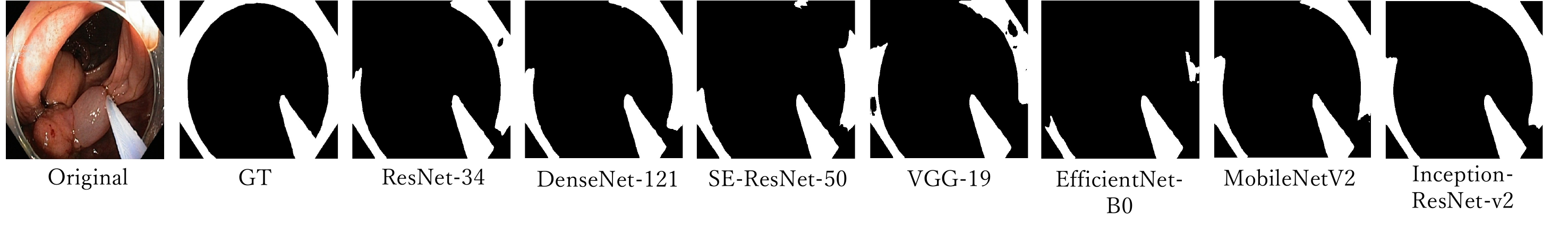}
\caption{Segmentation results comparison between different backbone networks (under segmented results)}
\label{fig:under_segmented}
\end{figure}

\begin{figure}[htb]
\includegraphics[width=\columnwidth]{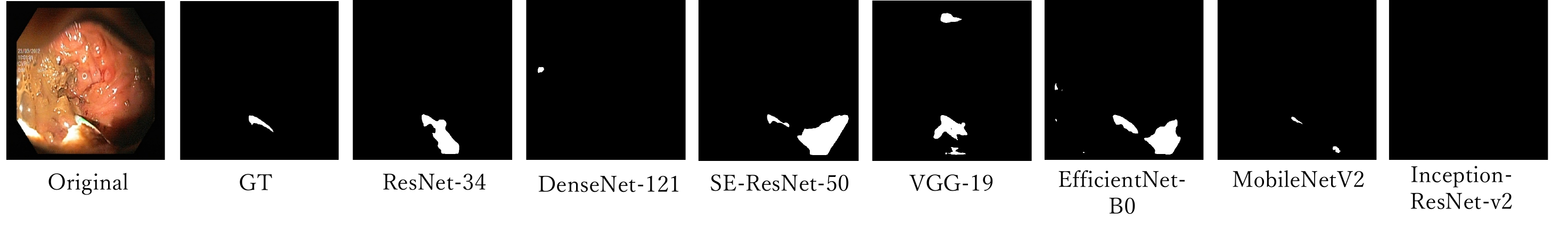}
\caption{Segmentation results comparison between different backbone networks (over segmented results)}
\label{fig:over_segmented}
\end{figure}

Figure \ref{fig:under_segmented} and \ref{fig:over_segmented} show the segmentation results for images considered as under-segmented (i.e., an excess in FN) and over-segmented (i.e., an excess in FP) by existing studies because of the effect of light-saturated regions.

\vspace{-2mm}
 \vspace{0.01mm}

 \vspace{-2mm}
 \vspace{0.01mm}

\vspace{-2.5mm}
\begin{landscape}
\begin{table}[htb]
\caption{Performance comparison between different backbone networks}
\label{Comparison of results for each method (U-Net)}
\begin{tabular}{llrrrrr}
\hline 
\multicolumn{1}{c}{\textbf{Method}}
& \multicolumn{1}{c}{\textbf{Backbone Network}}
& \multicolumn{1}{c}{\textbf{Jaccard C.}}
& \multicolumn{1}{c}{\textbf{DSC}}
& \multicolumn{1}{c}{\textbf{Precision}}
& \multicolumn{1}{c}{\textbf{Recall}}
& \multicolumn{1}{c}{\textbf{Acc.}}\\
\hline 
\hline 
U-Net & ResNet-34 & 0.9112 & 0.9490 & 0.9436 & \textbf{0.9633} & 0.9904\\
 & DenseNet-121 & 0.9008 &  0.9389 & 0.9353 & 0.9533 & 0.9896\\
  & SE-ResNet-50 & 0.8956 & 0.9339 & 0.9336 & 0.9533 & 0.9884\\
 & VGG-19 & 0.8969 & 0.9392 & 0.9318 & 0.9610 & 0.9878\\
 & EfficientNet-B0 & 0.9013 & 0.9416 &  0.9409 & 0.9556 & 0.9892\\
 & Inception-ResNet-v2 & \textbf{0.9167} & \textbf{0.9501} & \textbf{0.9543} & 0.9599 & \textbf{0.9912}\\
 & MobileNetV2 & 0.8953 & 0.9391 & 0.9289 & 0.9560 &  0.9887\\
\hline 
U-Net\cite{jha2021kvasir}& - & 0.8578 & 0.9158 & 0.8998 & 0.9487 & 0.9864\\
DoubleU-Net\cite{jha2021kvasir} & VGG-19 & 0.8430 & 0.9038 & 0.8966 & 0.9275 & 0.9838\\
\hline 
\end{tabular}
\end{table}
\end{landscape}

\subsection{Discussion}
The network using Inception-ResNet-v2 as the encoder of U-Net showed the highest values for the four metrics: JSC, DSC, precision, and accuracy.
The Inception-ResNet-v2 is a network that introduces the residual connection proposed by ResNet [10] to Inception-v3\cite{szegedy2016rethinking}, which is an evolution of the Inception module first proposed by GoogLeNett\cite{szegedy2015going}.
Inception is a small network structure consisting of convolutional layers of various sizes.
The 1$\times$1 convolution used in Inception reduces the number of parameters by reducing the dimensionality in the channel direction to prevent over-fitting. 
Several surgical instruments are achromatic (i.e., white or black) and have similar features in the channel direction.
Therefore, by including a 1$\times$1 convolution, the parameters were taught to capture shape features instead of the color features of the surgical instruments.
Artifact removal accuracy is an essential factor for preprocessing endoscopic images.
Thus, among the seven CNN architectures used in this experiment, Inception-ResNet-v2, which had the highest JSC value, is considered the most suitable architecture for the segmentation of surgical instruments in endoscopic images.

Figure \ref{fig:Study curve of training} and Figure \ref{fig:Study curve of validation} show that the value of Dice loss decreases at each epoch in both the training and validation sets and converges to approximately 0.
Therefore, we can expect that learning proceeds normally, and no over-fitting occurs. 

Figure \ref{fig:under_segmented} shows that the network using Inception-ResNet-v2 as the encoder of U-Net segmented a sufficient area for GT even for images considered under-segmented in existing research.
It can be confirmed that the network using Inception-ResNet-v2 can obtain good segmentation results compared to the GT even for under-segmented images.
Therefore, we believe that we have evaluated the performance of our architecture not only quantitatively but also qualitatively.
In this study, we used the U-Net as the basic architecture and compared it with existing research.

Furthermore, we believe that higher accuracy can be achieved for architectures with nonlinear convolution and spatial pyramid pooling, such as DeepLabV3+\cite{chen2018deeplabv3plus}.

\section{Conclusion}
In this study, we used the U-Net as the basic architecture, changed the encoder part to seven different CNN architectures to segment surgical instruments, and compared the results.
We observed that all the CNN architectures showed better results than those in previous studies, and the segmentation results differed depending on the architecture.
Among them, the network using Inception-ResNet-v2 showed the best performance with a JSC value of 91.7\%.
This result emphasizes the importance of the concept of model architecture, which is the basic architectural choice in obtaining good segmentation results.
In the CADx system, introducing a preprocessing method that properly removes artifacts from endoscopic images will improve the accuracy and reliability of the system.
In the future, it is necessary to incorporate the preprocessing method into the CADx system and confirm its impact on the diagnostic accuracy of the CADx system.




\bibliographystyle{unsrt}
\bibliography{AROB2022_twatanabe}

\end{document}